\documentclass[preprint,12pt]{elsarticle}
\usepackage{amsmath, amssymb, bm, textcomp}
\usepackage{listings, xcolor}
\usepackage{graphicx}
\usepackage{subfigure}
\usepackage{booktabs}
\usepackage{multirow}

\usepackage{float}

\usepackage[section]{placeins}

\journal{Radiation Detection Technology and Methods}
\date{May, 2024}

\begin{document}

\title{Monte Carlo Simulation of Angular Response of GRID Detectors for GRID Mission}

\author[add1]{Qize Liu}
\author[add1]{Xiaofan Pan}
\author[add1]{Xutao Zheng\corref{cor1}}
\ead{zhengxt18@mails.tsinghua.edu.cn}
\author[add1]{Huaizhong Gao}
\author[add1]{Longhao Li}
\author[add1]{Qidong Wang}
\author[add1]{Zirui Yang}
\author[add1]{Chenchong Tang}
\author[add1]{Wenxuan Wu}
\author[add1,add2]{Jianping Cheng}
\author[add1]{Zhi Zeng\corref{cor1}}
\ead{zengzhi@tsinghua.edu.cn}
\author[add1]{Ming Zeng}
\author[add3,add4]{Hua Feng}
\author[add5]{Binbin Zhang}
\author[add6]{Zhonghai Wang}
\author[add6]{Rong Zhou}
\author[add7]{Yuanyuan Liu}
\author[add8]{Lin Lin}
\author[add8]{Jiayong Zhong}
\author[add7]{Jianyong Jiang}
\author[add9]{Wentao Han}
\author[add1]{Yang Tian}
\author[add1]{Benda Xu}
\author[]{GRID Collaboration}

\cortext[cor1]{Corresponding authors. }
\address[add1]{Department of Engineering Physics, Tsinghua University, Beijing, 100084, China}
\address[add8]{Department of Astronomy, Beijing Normal University, Beijing, 100875, China}
\address[add2]{Beijing Normal University, Beijing, 100875, China}
\address[add3]{Institute of High Energy Physics, CAS, Beijing, 100049, China}
\address[add4]{Department of Astronomy, Tsinghua University, Beijing, 100084, China}
\address[add5]{School of Astronomy and Space Science, Nanjing University, Nanjing, 210093, China}
\address[add6]{College of Physics, Sichuan University, Chengdu, 610064, China}
\address[add7]{College of Nuclear Science and Technology, Beijing Normal University, Beijing, 100875, China}
\address[add9]{Department of Computer Science and Technology, Tsinghua University, Beijing, 100084, China}

\begin{abstract}

    The Gamma-Ray Integrated Detectors (GRID) are a space science mission that employs compact gamma-ray detectors mounted on NanoSats in low Earth orbit (LEO) to monitor the transient gamma-ray sky. 
    Owing to the unpredictability of the time and location of gamma-ray bursts (GRBs), obtaining the photon responses of gamma-ray detectors at various incident angles is important for the scientific analysis of GRB data captured by GRID detectors. 
    For this purpose, a dedicated Monte Carlo simulation framework has been developed for GRID detectors. 
    By simulating each GRID detector and the NanoSat carrying it, the spectral energy response, detection efficiency, and other angular responses of each detector for photons with different incident angles and energies can be obtained within this framework. 
    The accuracy of these simulations has been corroborated through on-ground calibration, and the derived angular responses have been successfully applied to the data analysis of recorded GRBs.
    
\end{abstract}

\maketitle

\section{Introduction}

The Gamma-Ray Integrated Detectors (GRID)~\cite{grid,gridnew} are a space mission concept aimed at monitoring transient gamma-ray phenomena using on-board compact gamma-ray detectors on NanoSats in low Earth orbit (LEO). 
The scientific objective of GRID is to collect gamma-ray transients, particularly gamma-ray bursts (GRBs) associated with the merger of two compact stars, as well as other gamma-ray transients, including long GRBs, soft gamma-ray repeaters, terrestrial gamma-ray flashes, and solar flares.

A constellation of NanoSats carrying at least 10 GRID detectors is projected to launch into LEO.
The first NanoSat was launched on 29 Oct 2018, and the first batch of scientific achievements obtained from GRID detectors has been published~\cite{ach1,ach2}. 
Currently, nine GRID detectors are operated on seven NanoSats operating in orbit for observation. 
The GRID detectors associated with the corresponding NanoSats processed according to the established methods are listed in Table~\ref{tab:Nano}.

\begin{table}[!htb]
\caption{GRID detectors' information in simulation framework}
\begin{center}
\begin{tabular}{lll} 
\toprule
Launch Date  & NORAD ID & Detector Name \\ 
\midrule
29 Oct 2018  & 43663 & GRID-01 \\
\midrule
6 Nov 2020 & 46838 & GRID-02  \\ 
\midrule
27 Feb 2022 & 51830 & GRID-03B/GRID-04 \\
\midrule
15 Jan 2023 & 55254 & GRID-05B  \\ 
\midrule
15 Jan 2023 & 55252 & GRID-06B  \\ 
\midrule
15 Jan 2023 & 55261 & GRID-07/GRID-08B \\
\midrule
22 June 2024 & 60088 & GRID-10B \\
\bottomrule
\end{tabular}
\label{tab:Nano}
\end{center}
\end{table}

This study focuses on the angular response of the GRID detector, which is introduced by the anisotropy of the NanoSat structure and the angular cross-section of the detector.
The detection efficiency and spectrum for photons of different energies and from different directions are obtained using the Monte Carlo method.
The currently launched GRID payloads all maintain a consistent payload structure and detector design. 
The payloads with a suffix of 'B' use FPGA-based data acquisition (DAQ) systems but still share a basic circuit board and detector structure with payloads without a suffix. 
For simplicity, the simulated results of the angular response results are discussed using the GRID-02 payload as an example.

The outline of this paper is as follows.
In Section 2, we describe the modelling and simulation method in detail.
In Section 3, the simulation results of the angular response are shown and analysed, and the verification of the simulation results based on on-ground calibration is also discussed.
We summarise the work completed in this paper and discuss its application in the GRID mission in Section 4.

\section{Methods}

\begin{figure}[!htb]
    \centering
    \includegraphics[width=0.9\textwidth]{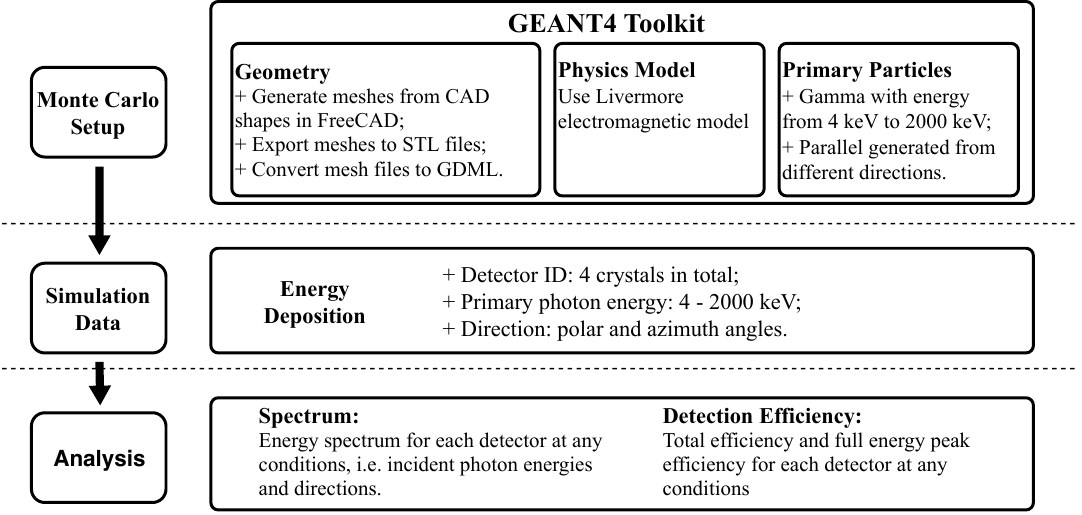}
    \caption{Flowchart of GRID angular response simulation}
    \label{fig:flowchart}
\end{figure}

The complete GRID angular response simulation procedure is presented in the flowchart (Figure~\ref{fig:flowchart}).
We use GEANT4 toolkit~\cite{geant4,g4,g4n} to develop the simulation software,
including the construction of the NanoSat geometry and materials,
the choice of physics interaction models and definition of incident photons as primary particles.
Because NanoSat was designed using computer-aided design (CAD) software~\cite{freecad}, conversion of the NanoSat's CAD model to GEANT4 geometry prevented us from re-modelling the entire satellite.
This was achieved by first generating meshes from the NanoSat components and then exporting and converting them to tessellated solids in GDML format~\cite{gdml}.
The energy deposition of each primary photon inside gadolinium aluminium gallium garnet (GAGG) crystals was recorded as the simulation output.
Analysis of these energy depositions and information of primary photons provided us with the energy spectrum and detection efficiency for any interacting conditions.

The following subsections introduce the structure of GRID detector as well as the entire NanoSat, the configuration of the Monte Carlo simulation, and the analysis method of the simulation results.

\subsection{GRID Detector}

A GRID detector is generally composed of four GAGG scintillators and accessory electronics. 
Different GRID models have been loaded onto the NanoSats and are operating in orbit. 

\begin{figure}[!htb]
    \centering
    \includegraphics[width=0.4\textwidth]{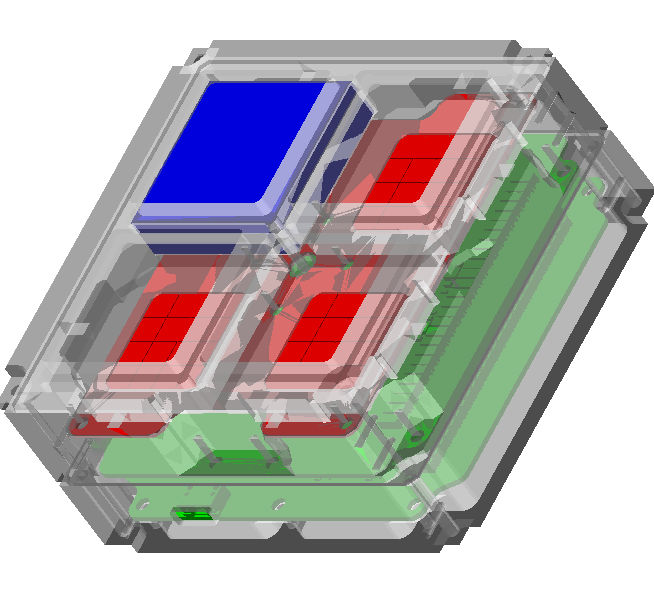}
    \caption{GRID-02 components}
    \label{fig:detector}
\end{figure}

Figure~\ref{fig:detector} shows the main components of the GRID detector.
In GRID-02, GAGG crystals of size 38 mm $\times$ 38 mm $\times$ 10 mm are covered using a reflective ESR film and placed at the top,
followed by silicon photomultipliers (SiPM) and their printed circuit boards (PCB).
The other two PCBs below the SiPM are the pre-amplifier and DAQ boards.
The entire structure is enclosed inside a case chassis made of aluminium alloy with a size of 94 mm $\times$ 94 mm $\times$ 50 mm.
Small components near the scintillators, such as screws, are also included in our geometry.

\begin{figure}[!htb]
    \centering
	\subfigure[]{
	\begin{minipage}{0.3\linewidth}
		\centering
		\includegraphics[width=0.9\linewidth]{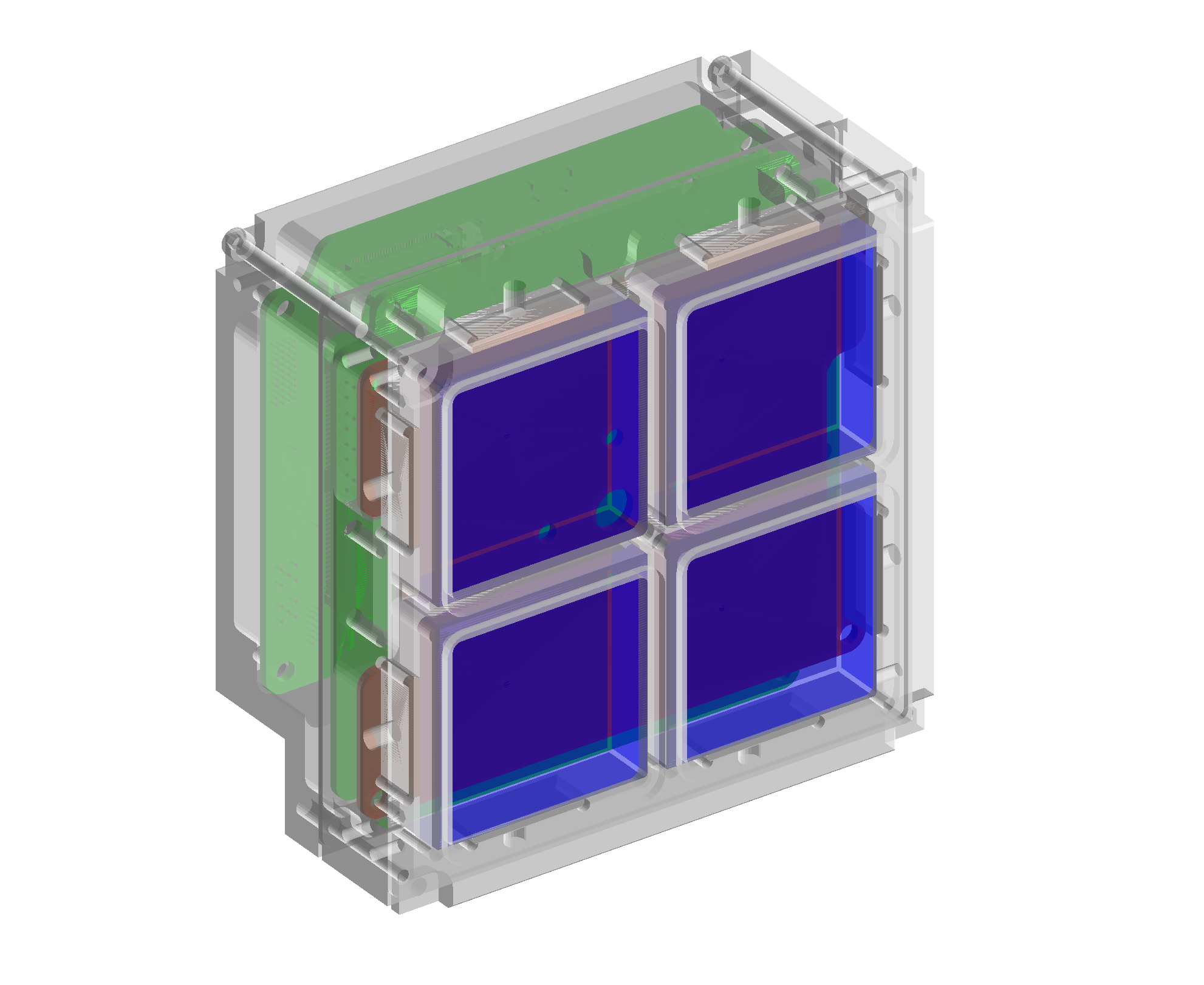}
    \end{minipage}}
    \subfigure[]{
    \begin{minipage}{0.3\linewidth}
        \centering
        \includegraphics[width=0.9\linewidth]{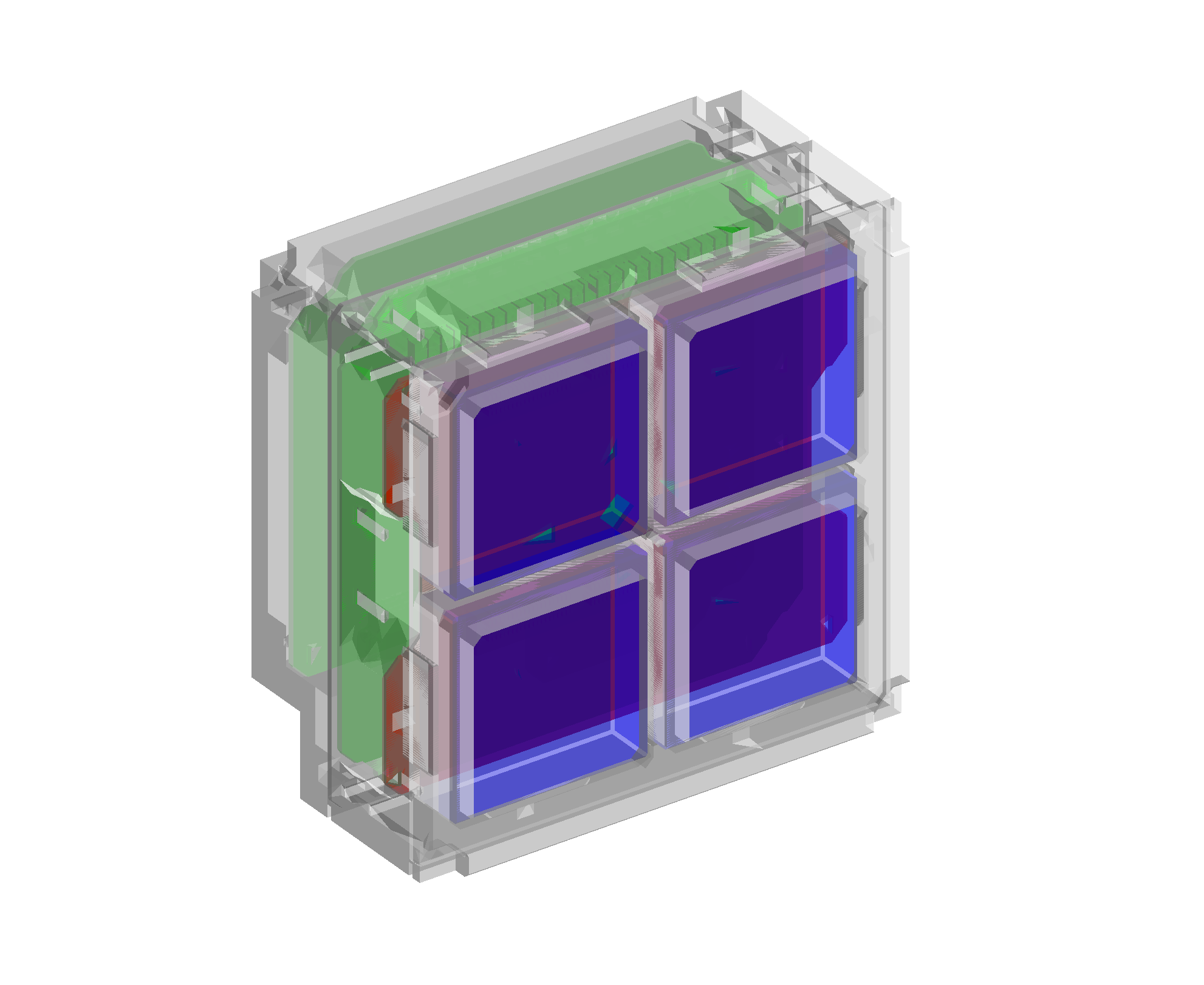}
    \end{minipage}}
    \subfigure[]{
    \begin{minipage}{0.32\linewidth}
        \centering
        \includegraphics[width=0.9\linewidth]{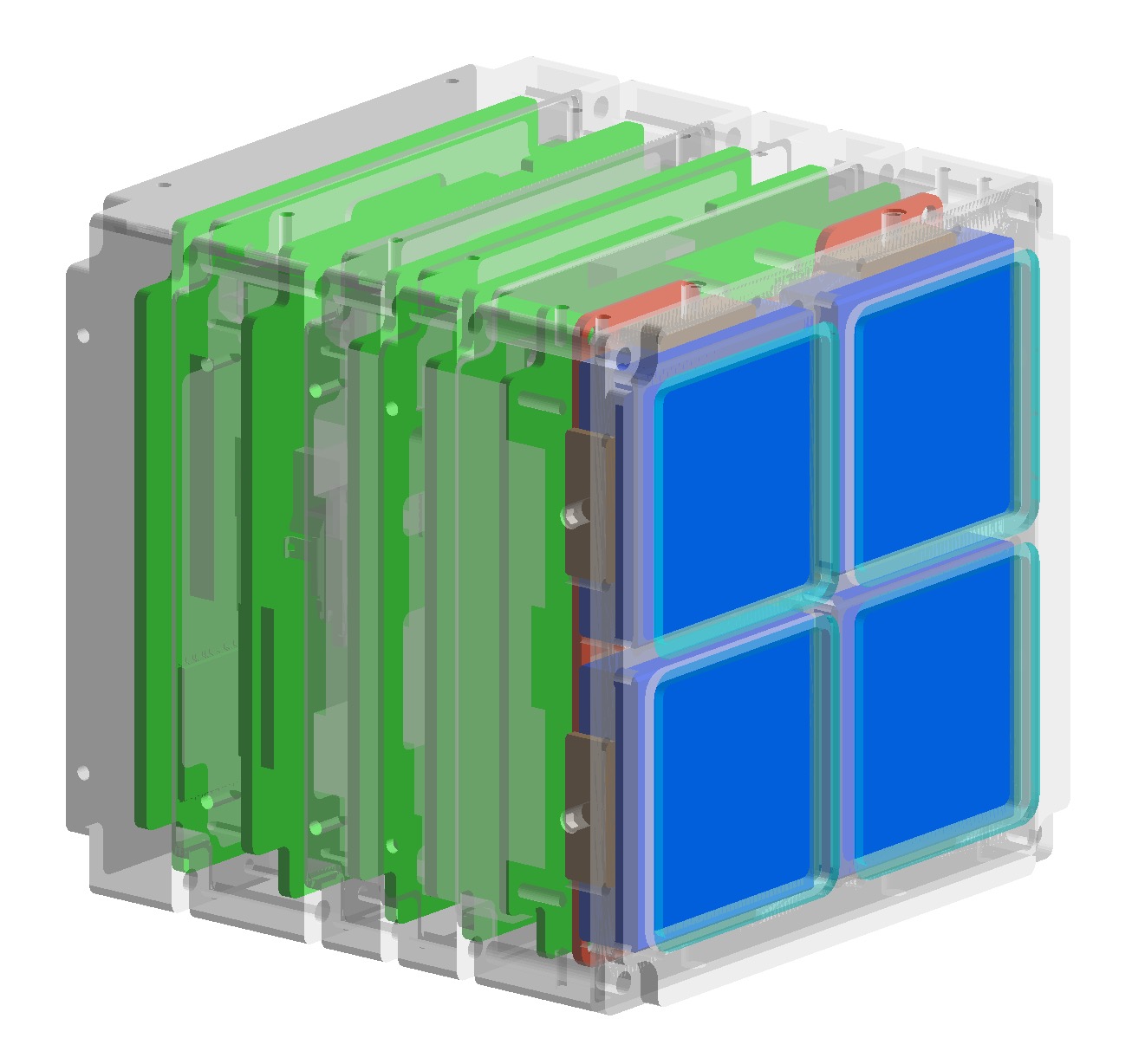}
    \end{minipage}}
    \caption{Models of other GRID detectors. (a) GRID-03B detector, (b) GRID-04 detector, (c) GRID-05B detector.}
    \label{fig:detectors}
\end{figure}

The modeling results of the other three designs of GRID detectors are shown in Figure~\ref{fig:detectors}. 
Each GRID detector has established a corresponding model based on its modification.

\subsection{NanoSat Structure}

\begin{figure}[!htb]
    \centering
    \includegraphics[width=0.8\textwidth]{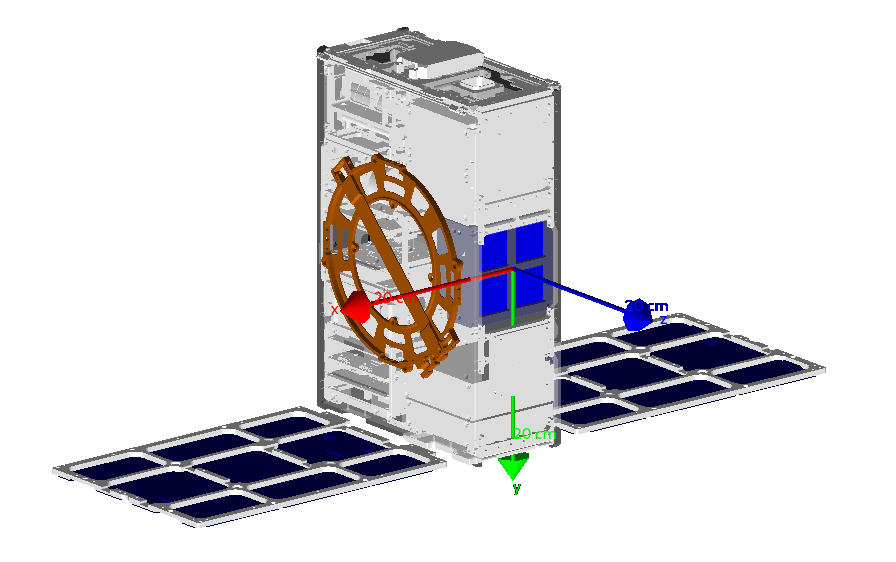}
    \caption{Simulation geometry of NanoSat equipped with GRID-02}
    \label{fig:cubesat}
\end{figure}

GRID-02 is part of the NanoSat made by SPACETY Company~\cite{spaceTY}.
Other devices are also present in the NanoSat.
For the Monte Carlo simulation, we have modelled several main modules in the geometry, including an X-ray polarimeter~\cite{polarlight}, communication module, control module, power module and solar panels.
All these components are supported by frames and panels.
The entire structure of the NanoSat equipped with the GRID-02 is shown in Figure~\ref{fig:cubesat}.

\begin{figure}[!htb]
\centering
\subfigure[]{
\begin{minipage}{0.45\linewidth}
    \centering
    \includegraphics[width=0.9\linewidth]{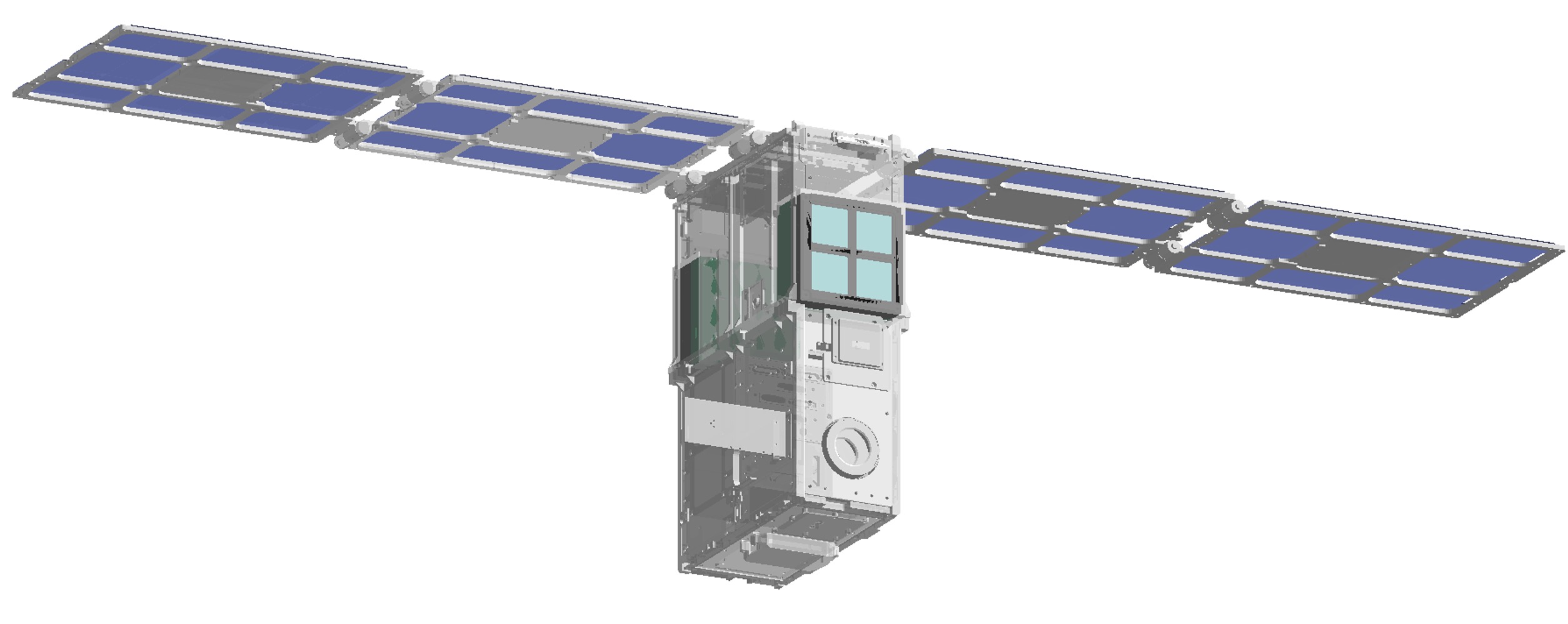}
\end{minipage}
}
\subfigure[]{
\begin{minipage}{0.45\linewidth}
    \centering
    \includegraphics[width=0.9\linewidth]{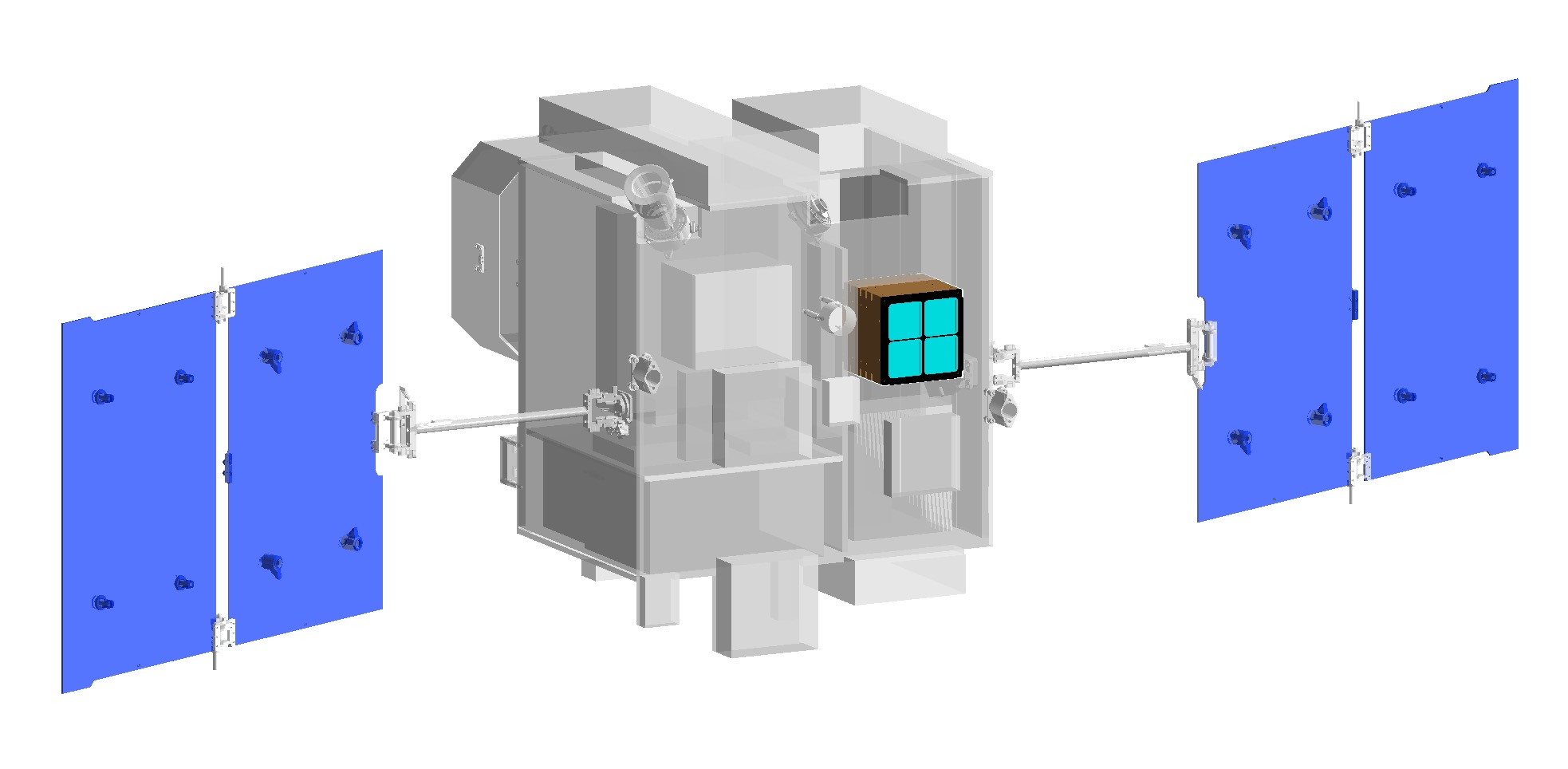}
\end{minipage}
}
\subfigure[]{
\begin{minipage}{0.8\linewidth}
    \centering
    \includegraphics[width=0.9\linewidth]{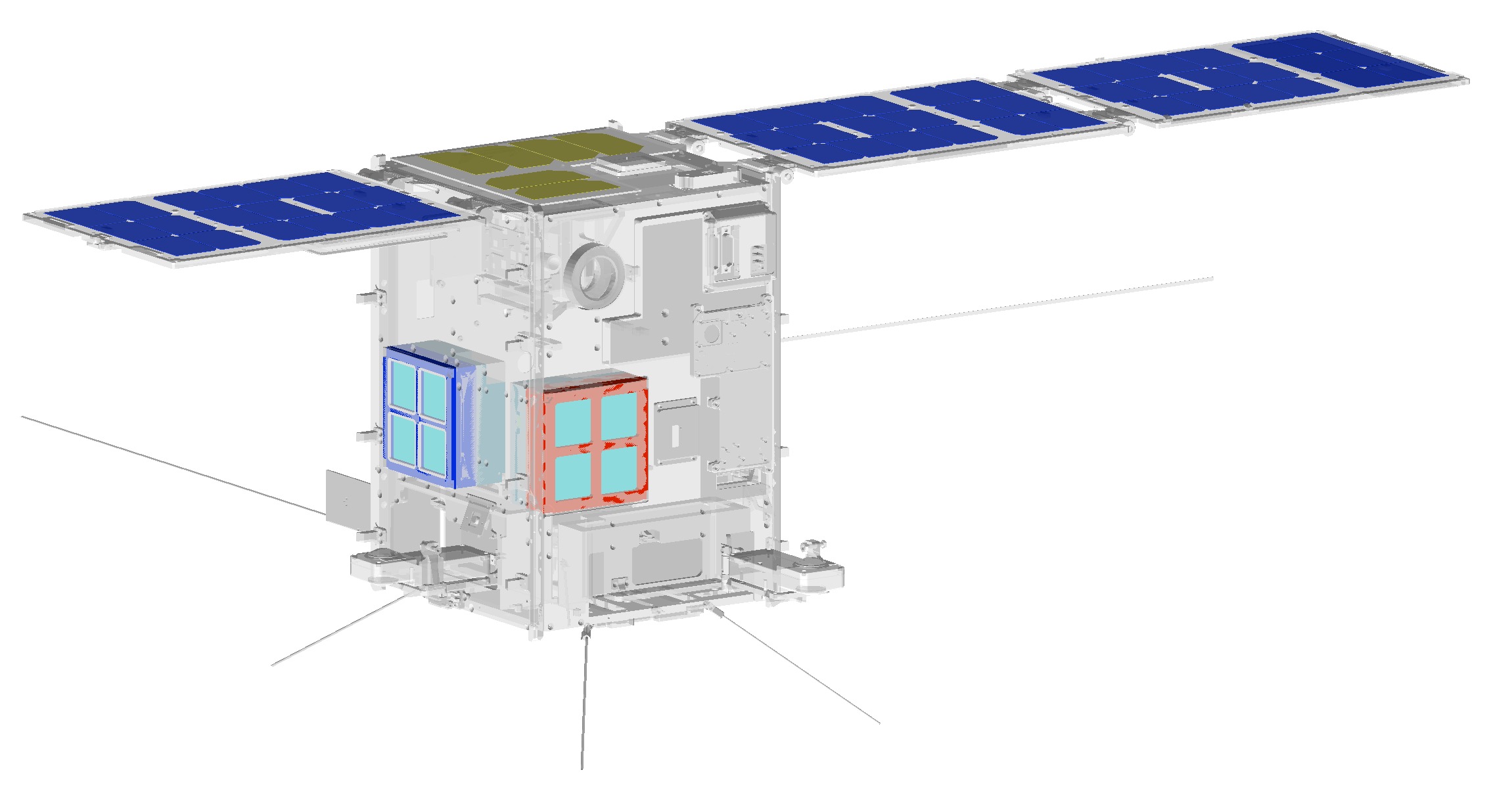}
\end{minipage}
}

    \caption{Models of other NanoSats. (a) NanoSat carrying GRID-03B and GRID-04, (b) NanoSat carrying GRID-05B, (c) NanoSat carrying GRID-07 and GRID-08B.}
    \label{fig:cubesats}
\end{figure}

GRID detectors with different models are installed on various NanoSats for operation. 
These NanoSats have undergone a modelling process similar to that of the NanoSat carrying GRID-02 to build their respective models. 
The modelling results of some other NanoSats are shown in Figure~\ref{fig:cubesats}.

\subsection{GEANT4 Configuration}

Materials should be assigned to the NanoSat geometry after converting the CAD files to GDML tessellated solids.
Table~\ref{tab:material} lists several important materials in the simulated geometry of the NanoSat carrying GRID-02.

\begin{table}[!htb]
	\centering
	\caption{Components of NanoSat carrying GRID-02}
	\label{tab:material}
	\begin{tabular}{ll}
		\toprule
        Component & Material  \\
        \midrule
        NanoSat mechanical structures & Aluminium alloy, fibreglass \\
        GRID detector chassis         & Aluminium alloy         \\
        GRID scintillators            & GAGG (1\% Ce doping)   \\
        SiPM                          & Silicon                \\
        PCB                           & FR4                    \\
        Solar panels                 & Silicon                \\
        Screws                        & Stainless steel        \\
        \bottomrule
	\end{tabular}
\end{table}

Because the photon energies of interest are between 4 and 2000 keV, we choose the Livermore low-energy electromagnetic model as the physics model to ensure the accuracy of the simulation.

Incident photons at our detector are approximately parallel to each other when GRBs occur far away from Earth.
Thus, the primary particles in the simulation are photons evenly distributed inside a plane and emitted parallelly.
The energy of the primary photons varies from 4 to 2000 keV for each simulation run.
We also vary the direction of the photons in each simulation by evenly selecting the directions on the spherical surface.
With variations in energy and direction for each simulation run, we can describe the angular response of the detector.

\subsection{Data Analysis}

From each run of the previously introduced Monte Carlo simulation, we obtain the energy depositions inside the GAGG crystals for each primary photon configuration, i.e.\ energy and direction.
The energy spectrum can be easily computed by assigning the energy depositions to a predefined histogram.
Detection efficiency is computed using Equation~\ref{eq:efficiency} as follows:

\begin{equation}
    \label{eq:efficiency}
    \eta = \dfrac{n}{\Phi} = \dfrac{n}{N_\text{sim} / S_\text{sim}}
\end{equation}
where $n$ is the number of detected photons, $N_\text{sim}$ is the total number of simulated primary photons and $S_\text{sim}$ is the area of the surface that generates the primary photons.

\section{Results}

\subsection{Angular Response of Efficiency}

\begin{figure}[!htb]
    \centering
    \includegraphics[width=\textwidth]{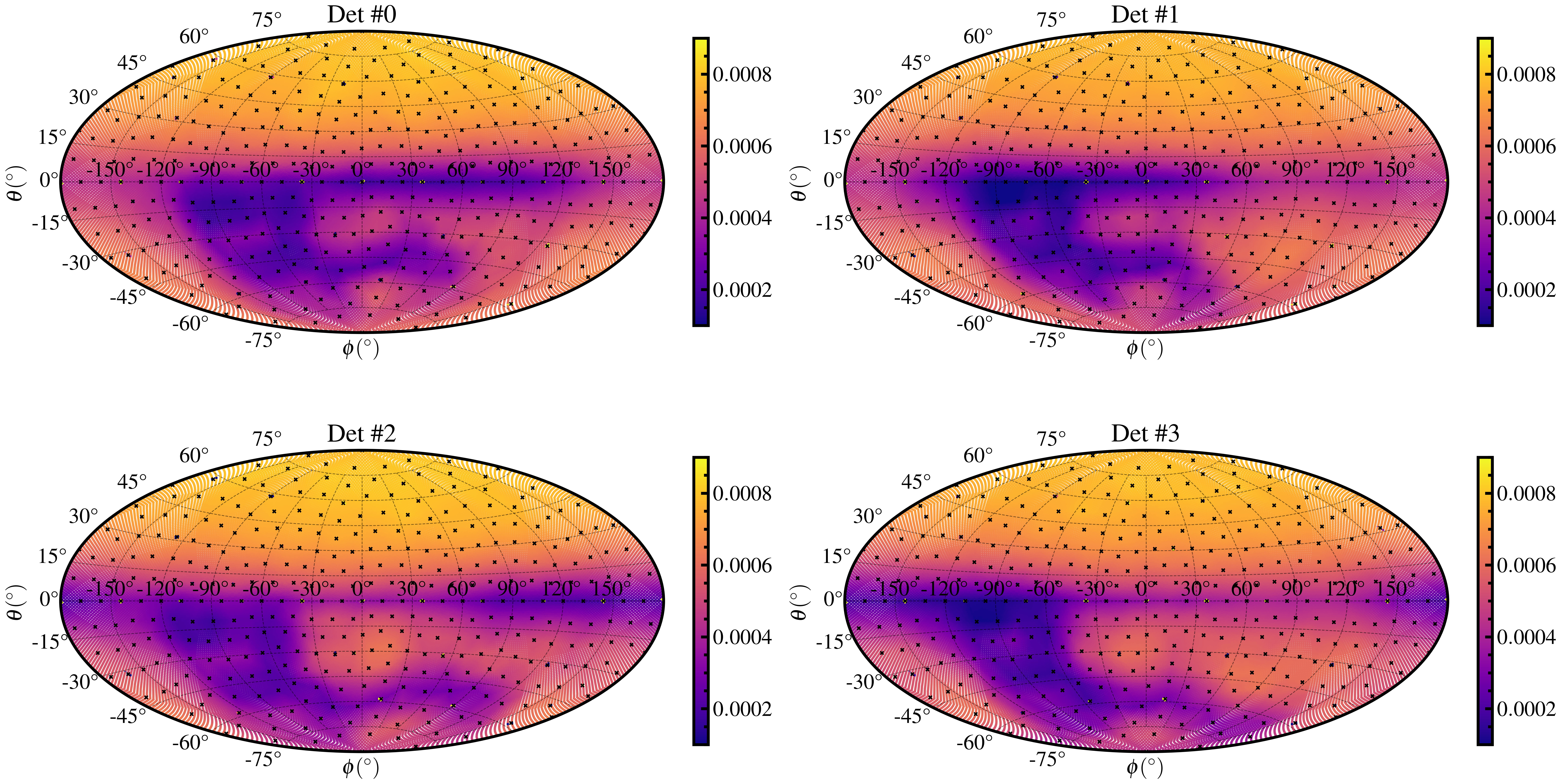}
    \caption{Detection efficiency for each direction (primary photon energy is uniformly distributed in the energy interval: 603.0-614.2 keV)}
    \label{fig:eff600}
\end{figure}

The angular response of each GAGG scintillator can be calculated as the detection efficiency for different  directions of the primary photons at a certain energy interval, as shown in Figure~\ref{fig:eff600}. 
The simulated directions are indicated by the black cross sign in Figure~\ref{fig:eff600}, and the efficiencies of the other directions are obtained by interpolating the three closest simulated directions.

The latitude of 90\textdegree{} corresponds to the photons coming from top of the GRID-02 and -90\textdegree{} indicates the photons coming from the bottom.
Four scintillators have different angular responses because of asymmetric shielding from the components of the NanoSat carrying GRID-02.

\begin{figure}[!htb]
    \centering
    \includegraphics[width=0.9\textwidth]{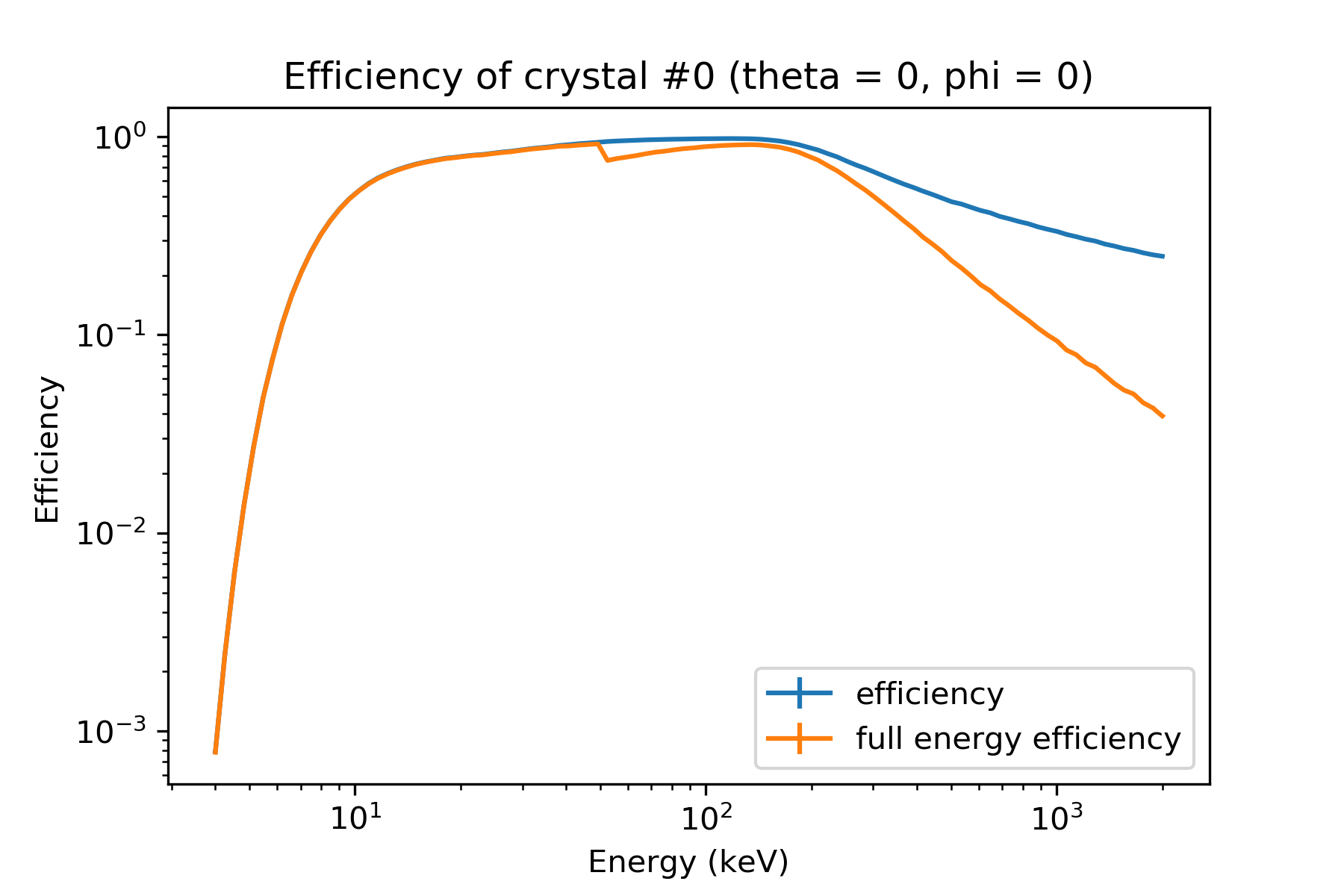}
    \caption{Efficiencies of photons emitted at different energy from the top}
    \label{fig:effcurve}
\end{figure}

A typical direction for incident photon is from vertically above, i.e.\ the latitude equals 90\textdegree{} in our definition.
For this direction, we have simulated 100 discrete energies of photons distributed in a square equal to the area of the GAGG crystal.
The efficiency versus energy curve is shown in Figure~\ref{fig:effcurve} from the results of this simulation.
Both total efficiency and full energy peak (FEP) efficiency are obtained by calculating the ratio of recorded events among total simulated photons.
In the FEP efficiency curve, an edge around 50 keV shows the effect of gadolinium K-shell electron absorption.

\subsection{Angular Response of Energy Spectrum}

\begin{figure}[!htb]
    \centering
\subfigure[]{
\begin{minipage}{0.9\linewidth}
    \centering
    \includegraphics[width=0.9\linewidth]{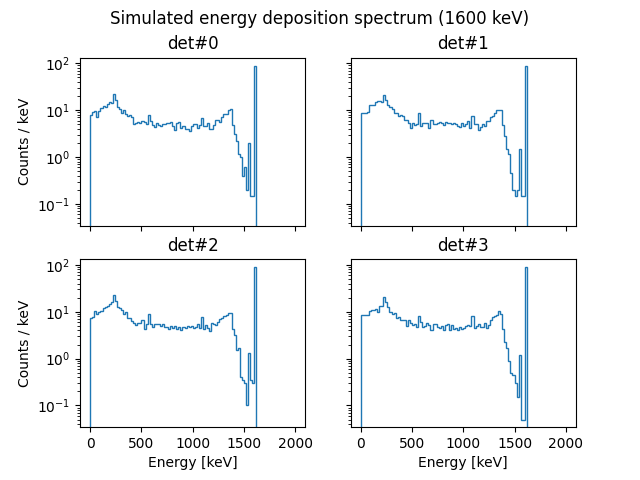}
\end{minipage}
}
\subfigure[]{
\begin{minipage}{0.9\linewidth}
    \centering
    \includegraphics[width=0.9\linewidth]{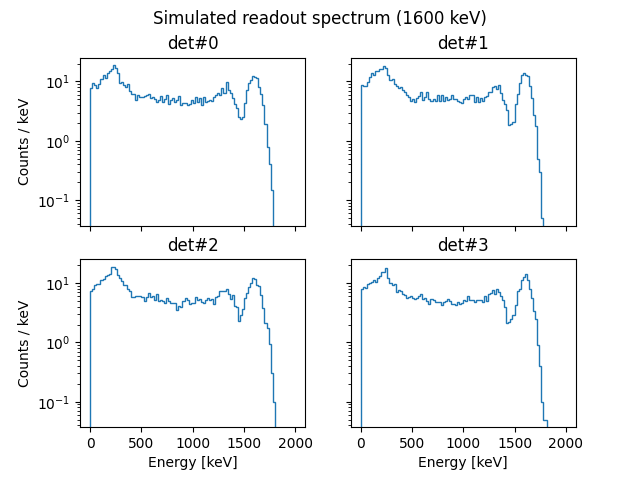}
\end{minipage}
}

    \caption{Energy spectrum for each scintillator (primary photon energy: 1600 keV). (a) Energy deposition in each scintillator, (b) Readout spectrum calculated from the simulation results and the energy resolution data obtained from experiments.}
    \label{fig:spec1600}
\end{figure}

As discussed in the last section, we can obtain the spectra from the energy depositions in GAGG scintillators for each simulation run.
Figure~\ref{fig:spec1600} shows the energy spectra in four scintillators corresponding to the 1600 keV incident gamma rays.
The scintillation and electronic responses of the GAGG detectors are not considered in the simulation; 
however, the readout spectrum can be calculated based on the simulation result and on-ground calibration results, 
considering that the amplitude of each signal generated from a certain energy deposition obeys the Gaussian distribution.

\subsection{Verification}

\begin{figure}[!htb]
    \centering
    \includegraphics[width=0.45\textwidth]{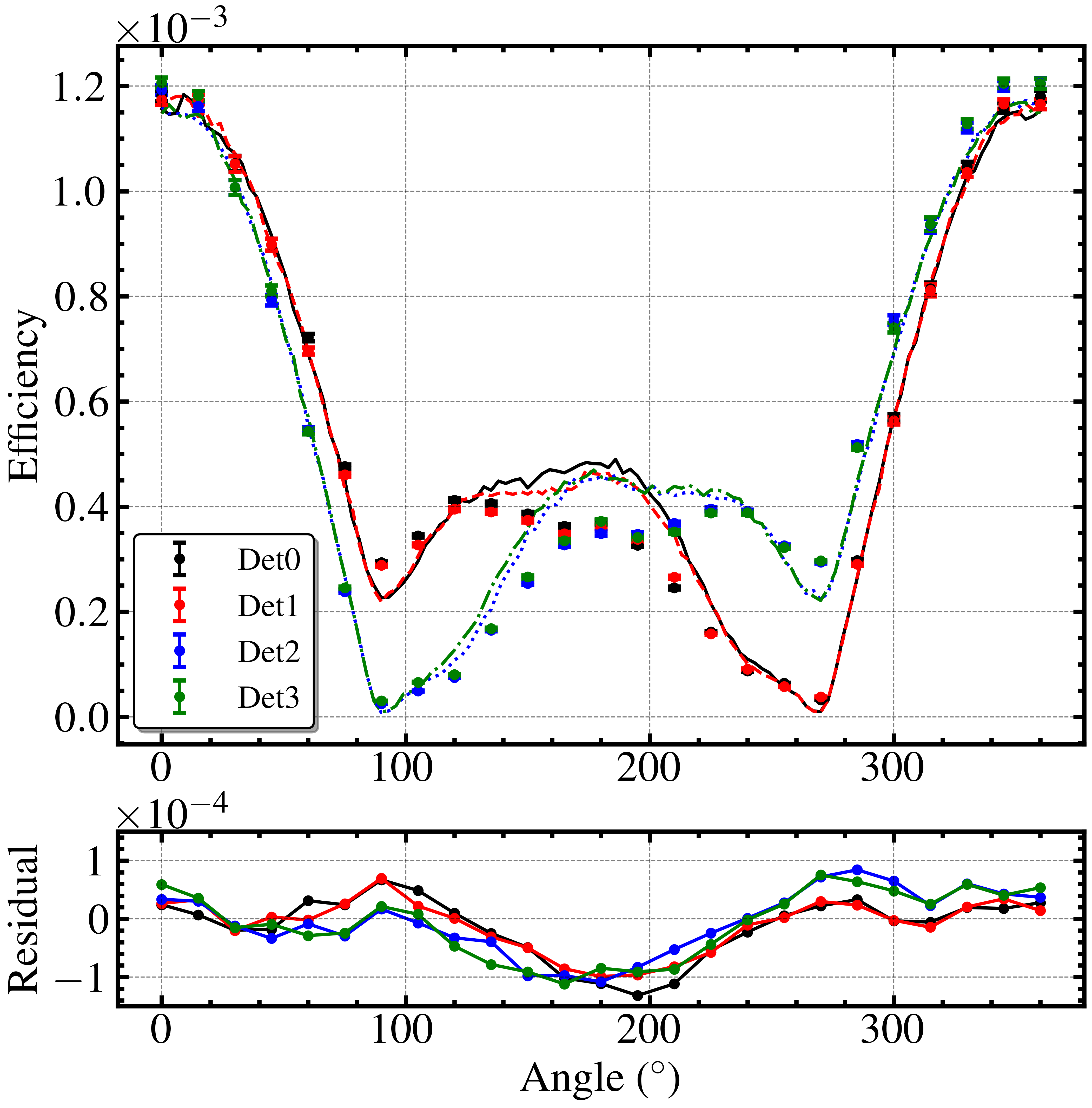}
    \includegraphics[width=0.45\textwidth]{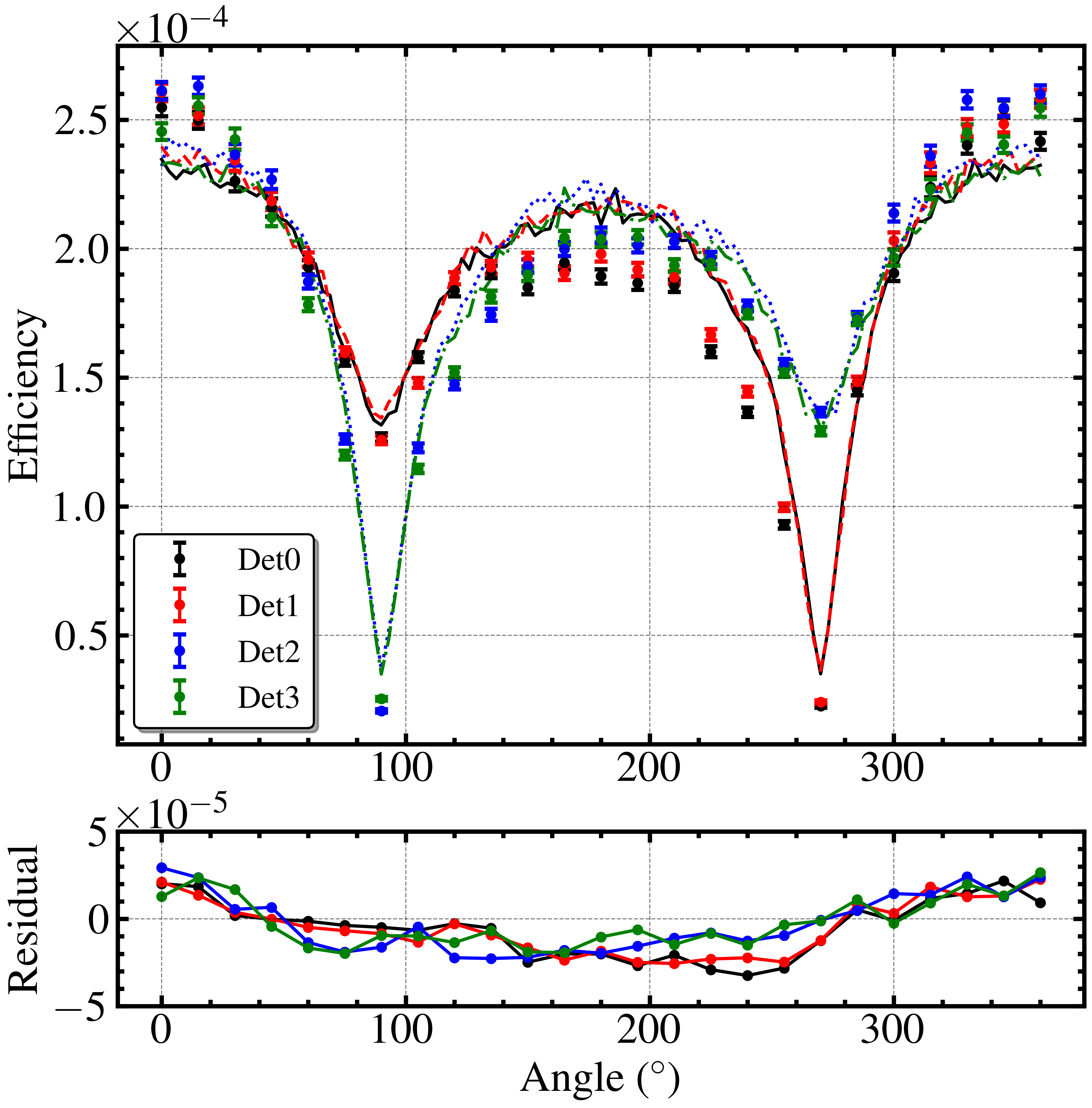}
    \caption{Measured (points) and simulated (curve) angular responses of the GRID-02. The responses are measured using  $^\text{241}$Am (left) and $^\text{137}$Cs (right) standard calibration sources. The measured data are taken from another study~\cite{gao}.}
    \label{fig:comp1}
\end{figure}

The simulation results have been verified by the calibration experiment.
The comparison between the simulation results for the angular response and efficiency of GRID-02 detector and the experimental calibration results is shown in Figure~\ref{fig:comp1}.
The experimental data are taken from the on-ground calibration experiment of GRID-02~\cite{gao}. 
The simulation results are consistent with the on-ground calibration experiment results of within an acceptable deviation range. 
The details of the on-ground calibration experiment are described in another study~\cite{gao}.

Before each GRID detector is launched, corresponding on-ground calibration experiments are performed, and the angular response of each GRID detector has been carefully verified through on-ground calibration experiments.

\section{Conclusion}

Herein, we employ the Monte Carlo method to simulate the angular responses of GRID detectors to photons of varying energies. 
A dedicated simulation framework tailored for the GRID mission has been established to allow the computation of the detection efficiency and the energy spectrum of incident photons. 
These parameters are crucial for the analysis of real data collected in orbit.

The modelling process from the engineering CAD file to the GEANT4 toolkit input is well-studied,
and has been applied to multiple GRID detectors and multiple NanoSat designs.
In addition, the properties of the incident photons, such as their energy range and directions, can be easily adjusted to accommodate different simulation jobs.

For the characterisation of the angular response of the GRID detector, the simulation result has undergone complete verification through on-ground calibration experiments. 
The specific angle response of the GRID detector has been applied to the scientific analysis of GRB 210121A~\cite{ach1}, GRB 220408B~\cite{GRBevt} and other observed GRB events.

\section*{Acknowledgement}

This work was supported by the Tsinghua University Initiative Scientific Research Program and the Tsinghua-Kunshan Student Innovation and Entrepreneurship Talent Cultivation Agreement. Thanks to the contribution of Weihe Zeng, Ze She and Xingyu Pan for their help in geometry conversion and Monte Carlo simulation. Thanks to the contribution of Hong Li for her help in data processing.

\section*{Conflict of Interest Statement}
On behalf of all authors, the corresponding author states that there is no conflict of interest.

 \bibliographystyle{elsarticle-num} 
 \bibliography{cas-refs}

\end{document}